\documentclass[runningheads]{svmult}

\usepackage{makeidx}   
\usepackage{graphicx}  
\usepackage{subeqnar}  
\usepackage{multicol}  
\usepackage{physprbb}  
\makeindex             

\begin{document}
\title*{Probability in Orthodox Quantum Mechanics:\protect\newline 
Probability as a Postulate\protect\newline
~~~~~ Versus \protect\newline
Probability as an Emergent Phenomenon}
%
%
%
%
\titlerunning{Probability in Quantum Mechanics}
%
\author{Stephen L. Adler}

\authorrunning{Stephen L. Adler}
%
%
\institute{Institute for Advanced Study, Einstein Drive, Princeton NJ 08540, USA}

\maketitle              

\begin{abstract}
The role of probability in quantum mechanics is reviewed, 
with a discussion of the ``orthodox'' versus the statistical 
interpretative frameworks, and of a number of related issues.  After a 
brief summary of sources of unease with quantum mechanics, a survey 
is given of attempts  either to give a new interpretive framework 
assuming quantum mechanics is exact, or to modify quantum mechanics 
assuming it is a very accurate approximation to a more fundamental 
theory.  This survey focuses particularly 
on the issue of whether probabilites in quantum 
mechanics are postulated or emergent. 
\end{abstract}

\section{Orthodox Quantum Mechanics and Issues it Raises}

Quantum mechanics (QM) is our most successful physical theory, 
encompassing phenomena as diverse as chemical bonding, the band 
structure of solids, and the standard model 
of particle physics.  But the probabilistic aspect of quantum mechanics 
has been a source of unease from the outset.  As surveyed
by Home [1], the Founding Fathers were divided over this 
aspect of the theory that they had created:  Bohr, Born, and 
Heisenberg were comfortable with the probabilistic structure of 
quantum theory, whereas Einstein and Schr\"odinger had profound 
reservations.  This unease, and division, have continued to the present. \subsection{Postulates of Quantum Mechanics} 
Let us begin with a review of the postulates of QM, in the arena 
of a complex Hilbert space,  following for 
the larger part the presentation of Ballentine [2].
\begin{itemize}
\item Observables are associated with self-adjoint operators.
Thus we have
\begin{eqnarray}
S=\sum_ns_n P_n~~~,\\
P_n=\sum_a|a,s_n\rangle \langle a,s_n|~~~,
\end{eqnarray}
with $S$ an operator, $s_n$ its eigenvalues, $P_n$ the corresponding 
orthogonal projectors, and $a$ a label that distinguishes degenerate 
eigenvectors.
\item Each state is associated with a density matrix $\rho$, which 
is self-adjoint, nonnegative, and has trace unity, 
\begin{eqnarray}
\rho=\rho^{\dagger}~,\\
\rho \geq 0~~,~~~{\rm Tr}\rho=1~~,
\end{eqnarray}
so that the spectral decomposition of $\rho$ takes the form
\begin{eqnarray}
\rho=\sum_n \rho_n |\phi_n\rangle \langle \phi_n|~,\\
0 \leq \rho_n \leq 1~,~~~\sum_n \rho_n=1~~~.
\end{eqnarray}
\item A pure state is asssociated with an idempotent density matrix,
satisfying $\rho^2=\rho$.  This condition, together with the condition 
of unit trace, implies that there is exactly one nonzero eigenvalue 
$\rho_n=1$, with all other eigenvalues $\rho_{n^{\prime}}~,n^{\prime}
\not= n$ vanishing, and so for a pure state the spectral decomposition 
or the density matrix reduces to 
\begin{eqnarray}
\rho=|\phi_n\rangle \langle\phi_n|~~~.
\end{eqnarray}
\item  The average of an observable $S$ in a general state $\rho$ 
is given by 
\begin{eqnarray}
\langle S \rangle = {\rm Tr} \rho S~~~.
\end{eqnarray}
In particular, for a pure state with $\rho=|\psi \rangle \langle \psi |~,~~~\langle \psi|\psi \rangle=1$, we have 
\begin{eqnarray}
\langle S \rangle = \langle \psi|S|\psi \rangle~~~.
\end{eqnarray}
\item An observable $S$ can only take the values $s_n$ of its 
eigenvalues.  The \emph{probability} of finding the eigenvalue $s_n$ 
in a normalized pure state $|\psi_n\rangle$ is 
\begin{eqnarray}
p_n=\sum_a |\langle \psi|a, s_n \rangle|^2~~~.
\end{eqnarray}
\item  Within coherent sectors of Hilbert space, superpositions of pure 
states are pure states, and self-adjoint compositions of observables 
are observables.
\item  The dynamics of the density matrix $\rho$ and of a state 
$\psi$ in Hilbert space is specified by
\begin{eqnarray}
\rho(t)=U\rho(t_0)U^{-1}~~~,\\
|\psi(t)\rangle=U|\psi(t_0)\rangle~~~.
\end{eqnarray}
Here $U(t,t_0)$ is a unitary operator, which for small $t-t_0=\delta t$ 
takes the form 
\begin{eqnarray}
U=\rm{exp}\left(-{\I \over \hbar} H(t) \delta t\right)~~~,
\end{eqnarray}
defining the (possibly time dependent) system Hamiltonian.  This 
dynamics is termed the ``U operation'' by Penrose [3].
\item  We finally come to the effect of a measurement.  After the 
measurement of the eigenvalue $s_n$ in a pure state, the new system 
state is 
\begin{eqnarray}
|\psi^{\prime} \rangle = { \sum_a |a,s_n \rangle 
\langle a,s_n|\psi\rangle  \over
[\sum_a |\langle a,s_n |\psi\rangle|^2 ]^{1\over 2} }~~~.
\end{eqnarray}
This  equation summarizes what is termed 
the ``R operation'' by Penrose.  
\end{itemize}
\subsection{Interpretive Framework}
While everyone agrees that the above postulates provide a practical 
set of rules for making predictions in quantum mechanics, and that 
these predictions to date have always agreed with experiment, there 
is a dichotomy when it comes to giving an interpretive framework 
for the rules.  
\begin{itemize}
\item  On the one hand, we have the ``orthodox'' interpretation, 
as given, for example, in the text of Mandl [4].  This asserts that 
the state $|\psi \rangle$ gives a complete description of an 
\emph{individual} system, and that (14) corresponds to ``reduction''
of the individual's state vector.
\item  On the other hand, we have the ``statistical'' interpretation, as 
discuused, for example, in the review of Ballentine [2], according to 
which the state $|\psi\rangle$ describes certain statistical properties 
of an \emph{ensemble} of similarly prepared systems.  According to 
this interpretation, (14) corresponds to the preparation of a new 
ensemble by a measurement.  There \emph{may} be, or there \emph{may not} 
be, hidden variables that specify a complete, nonstatistical 
interpretation of individual systems:  the statistical interpretation 
is agnostic with respct to this issue.
\end{itemize}
\subsection{Why the ``R'' Operation is Needed}
To see why the ``R'' operation is needed, let us demonstrate that 
the measurement process cannot be represented by a \emph{deterministic},
\emph{unitary} evolution on a closed system.  Let us consider a 
Stern-Gerlach experiment, in which an initial spin eigenstate 
$|\psi \rangle$ with 
eigenvalue $1/2$ along the $x$ axis is separated, by means of a magnetic 
field that is inhomogeneous along the $z$ axis, into orthonormal 
states $|\psi_{\uparrow}\rangle~,~~|\psi_{\downarrow}\rangle$ that 
have respective spin eigenvalues $1/2, -1/2$ along the $z$ axis.  Thus, 
at the detector one sees either $|\psi_{\uparrow}\rangle$ or 
$|\psi_{\downarrow}\rangle $, with 
\begin{eqnarray}
\langle \psi_{\downarrow}|\psi_{\uparrow}\rangle=0~~~.
\end{eqnarray}
Let us suppose that these final states evolved from the initial state 
by the deterministic unitary ``U''process, which would imply that 
\begin{eqnarray}
|\psi_{\uparrow}\rangle = U |\psi \rangle~,\\
|\psi_{\downarrow}\rangle = U |\psi \rangle~~~.
\end{eqnarray}
This would imply
\begin{eqnarray}
0=\langle \psi_{\downarrow}|\psi_{\uparrow} \rangle
=\langle \psi| U^{\dagger}U|\psi \rangle=\langle \psi | \psi \rangle=1
~~~,
\end{eqnarray}
which is a contradiction.  Hence, the measurement process, in which 
an initial coherent superposition of states leads to a definite but 
unpredictable outcome, cannot be described by a deterministic unitary 
time evolution.  Thus measurement involves a physical process, which 
we have called the ``R'' process, that is distinct from the 
deterministic unitary ``U'' process that governs the unobserved 
evolution of the quantum system.  

However, the ``R'' process does not have to be nonunitary; one can 
have for the $i$th atom going through the apparatus the evolution 
\begin{eqnarray}
|\psi_i\rangle=U_i|\psi\rangle~~~,
\end{eqnarray}
with $U_i$ a unitary evolution that is different for each $i$.  
This is possible 
because any path through Hilbert space connecting two normalized pure 
states can be described by a succession of infinitesimal unitary 
transformations.  To prove this, it suffices to consider the 
infinitesimal segment $|\psi\rangle \to 
|\psi\rangle + |\D \psi \rangle$, with $\langle \psi | \D \psi \rangle
=0$.  If we take
\begin{eqnarray}
U=1+|\D \psi \rangle \langle \psi|-|\psi \rangle \langle \D \psi |~~~,\\
U^{\dagger}=1-|\D \psi \rangle \langle \psi|+
|\psi \rangle \langle \D \psi |~~~,
\end{eqnarray}
then we have $U^{\dagger}U=UU^{\dagger}=1$ up to an error of second 
order, and $U|\psi \rangle=|\psi\rangle + |\D \psi \rangle$, as needed.  
So it is perfectly possible for the ``R'' process to be described by 
a \emph{stochastic} unitary process, constructed from a sequence 
of random or partially random infinitesimal unitary transformations,   
and we will see in Sect. 3.1  examples of how this can be accomplished.

\subsection{Micro Versus Macro}

In QM, the probability is the squared modulus of the probability 
amplitude.  Probability  amplitudes superimpose coherently, and between 
measurements evolve in time by the ``U'' process.    
Thus, in the microscopic realm:
\begin{itemize}
\item One sees coherent superpositions.
\item Amplitudes evolve through deterministic, unitary evolution.
\end{itemize}

On the other hand, in the macroscopic realm:
\begin{itemize}
\item One does \emph{not} see coherent superpositions, e.g., of dead 
and alive cats. 
\item Measurements involve the ``R'' process, which is not deterministic 
unitary.
\end{itemize} 

This dichotomy leaves us with the following questions (for 
more detailed discussions, see Penrose [3] and Leggett [5]):
\begin{itemize}
\item  Where is the dividing line between ``micro'' and ``macro''?
\item  What is responsible for it?
\end{itemize}

\subsection{Postulated Versus Emergent Probability}

A unique feature of orthodox QM is that it is the only probabilistic 
theory where the probabilities are \emph{postulated} ab initio, and are 
not \emph{emergent} from unobserved, deterministic phenomena at a deeper 
level.  A typical theory where probabilites are emergent is statistical 
mechanics.  In statistical mechanics one starts from a probability 
postulate of uniform phase space occupation.  This assumption, and 
the related concept of an equilibrium ensemble ( which reflects the 
implications of conserved quantities for the phase space occupation), 
is consistent because of Liouville's theorem, which implies that a 
uniform distribution is preserved in time by the dynamics.  

However, these probabilistic statements are not the end of the story 
in statistical mechanics.  There are underlying laws -- the equations 
of classical molecular dynamics -- which are deterministic; no 
probabilities enter into their formulation.  These laws lead, by a 
process that is still not completely understood (as reflected in 
discussions of ergodicity at this Conference), to an effectively uniform 
phase space distribution for systems that are sufficiently complex.  
Thus, the probabilistic theory of statistical mechanics 
is \emph{emergent} 
from the deterministic theory of classical mechanics.

\subsection{Recapitulation}

To sum up, there are a number of sources of unease about QM:
\begin{itemize}
\item  There is no predictive description of individuals. 
\item There is a micro-macro divide of unclear origin.  
\item  There is a probabilistic structure that is postulated rather 
than emergent.
\end{itemize}

But QM works!  Many subtle and remarkable predictions of QM are 
experimentally verified in many different physical arenas.  Thus either
\begin{itemize}
\item  (A) QM is exact, but to deal with the sources of unease it needs 
reinterpretation at the conceptual level (although no modification 
of the standard postulates is needed to use QM  as a practical 
computational and predictive tool).  

\item  (B) QM is not exact, but rather is a very accurate asymptotic 
approximation to a deeper level theory.
\end{itemize}
I do not believe that it is just a matter of taste which of these 
possibilites is chosen, because the distinction between (A) and (B) 
is relevant to the issue of Planck scale unification of the forces 
and the particle spectrum.  If QM changes, it may profoundly influence 
the ground rules for unifying gravity with the other interactions.  

\section{Reinterpretations of Quantum Mechanics  Assuming it is Exact}
Let us now review four differing approaches based on premise (A), 
that QM is exactly correct but in need of reinterpretation.  Our focus 
in each case will be on the extent to which the probabilistic structure 
is postulated or is emergent.  
\subsection{Everett's ``Many Worlds'' Interpretation}  

In the ``Many Worlds'' interpretation introduced by Everett [6] and 
discussed in further detail in the articles collected in [7], 
there is no 
state vector reduction; instead, there is only Schr\"odinger evolution 
of the wave function of the entire universe.  To describe $N$ successive 
measurements in this interpretation requires an $N$-fold tensor 
product wave function.  

Probability is not emergent, but rather is postulated in the Everett 
picture.  Everett introduces a measure on the coefficients of the 
final superposition resulting from $N$ successive measurements, which as 
$N \to \infty$ behaves mathematically 
like the usual QM probability rule.  There is 
a logical jump (or an implicit assumption -- this is still a matter of 
debate) in going from 
 the Everett measure on tensor product coefficients 
to statements about the relative frequencies of experimental outcomes.   

\subsection{The Histories Approach}

The so-called ``Histories'' approach has been extensively developed 
recently by Griffiths, Omn\`es, and Gell-Mann and Hartle (for a review 
and references see [8], and for a semipopular account see [9]).  
The histories approach takes as a given that QM is a stochastic theory; 
probability is introduced as a postulate, and is not emergent.  The 
basic objects in the histories approach are time-dependent projectors 
$E_k(t_k)$ associated with properties occurring in a history.  The 
probability of a history is then postulated to be given by 
\begin{eqnarray}
p={\rm Tr}[E_n(t_n)...E_1(t_1)\rho E_1(t_1)...E_n(t_n)]~~~,
\end{eqnarray}
with $\rho$ the density matrix at the intial time.  This definition 
can be shown to lead, under appropriate circumstances, to all of the 
expected properties of probabilities.  In this interpretation, state 
vector reduction appears only in the statistical interpretation sense 
discussed above, as a rule for relating the density matrix after a 
measurement to the density matrix before the measurement.
\begin{eqnarray}
\nonumber
\end{eqnarray}
In both the ``Many Worlds'' and the ``Histories'' interpretations, 
there is by definition no concept of the ``individual''.  We shall 
now discuss two other currently studied interpretations of QM that 
enlarge the mathematical framework to give an ``individual''. 

\subsection{Bohmian Mechanics}
There has been a recent revival of interest in Bohmian mechanics 
(see [10] for a technical account and references, and [11] for a 
semipopular account).  In Bohmian mechanics, in addition to the 
Schr\"odinger equation for the $N$-body wave function 
$\psi(\vec q_1,..., \vec q_N,t)$, 
\begin{eqnarray}
\I \hbar {\partial \psi \over \partial t} 
=-\sum_{k=1}^N {\hbar^2 \over 2 m_k} \nabla^2_{\vec q_k} \psi 
+V \psi~~~,
\end{eqnarray}
the mathematical framework is enlarged by introducing  hidden  
``particles'' moving in configuration space with coordinates 
$\vec Q_k$ and velocities 
\begin{eqnarray}
\vec v_k={\D \vec Q_k \over \D t} ={ \hbar \over m_k} {\rm Im} 
\nabla_{\vec Q_k}{\rm log} \psi(\vec Q_1,...,\vec Q_k,t)~~~.
\end{eqnarray}
The state of the 
``individual'' is then specified by giving both the wave function 
\emph{and} the coordinates $\vec Q_k$ of the hidden particles.  A probability 
postulate is introduced, that the probability distribution on configuration space obeys $p=|\psi|^2$ at some initial time $t_0$.  The Bohmian 
equations given above then imply that this remains true for all 
times subsequent to $t_0$; the logic here resembles the use of the 
Liouville theorem in statistical mechanics.  Unlike statistical 
mechanics, Bohmian mechanics 
has no underlying molecular dynamics-like layer, so the probabilites 
are not prima facie emergent.  We  note, however, 
 that in [10] arguments are given 
(and are further discussed at this Conference) that the Bohmian 
probability postulate follows from considerations of ``typicality'' of 
initial configurations 
(in distinction to the ergodicity arguments used in attempts to derive 
the postulates of statistical mechanics from the equations of molecular 
dynamics).  

\subsection{The Ax-Kochen Proposal}

Recently Ax and Kochen [12] have extended the mathematical framework 
of QM in a different way to encompass the ``individual''.  They identify 
the \emph{ray} with the ensemble, and the \emph{ray representative}, 
i.e. the $U(1)$ phase associated with a particular state vector, with 
the individual.  They then give a mathematical construction to specify 
a unique physical state from knowledge of 
the toroid of phases. They introduce a probability 
assumption, that the a priori distribution of phases is uniform, and 
then show that, by their construction, this implies that 
the probabilities of outcomes obey the usual QM rule.  Thus, 
probability in the Ax-Kochen interpretation is not emergent, but 
their probabilistic postulate is arguably weaker than that in standard 
QM or in Bohmian QM.

\subsection{Are Interpretations of Quantum Mechanics Falsifiable?}
We conclude this brief survey of alternative interpretations of 
an assumed exact quantum mechanics by posing the question, can the 
interpretations given above be falsified?  By construction, 
the four interpretations described above are designed to agree with 
the predictions of standard QM.  Clearly, if an interpretation  
could be shown to \emph{differ} in some prediction from that 
of QM, and if this difference in predictions were resolved 
experimentally in favor of standard QM (in the way that the Bell 
inequalities have been tested and favor QM), then the interpretation 
would be falsified.  But suppose that an interpretation makes 
empirical predictions that, within the domain in which the rules of 
QM apply, are without exception indistinguishable from the predictions 
of QM.  Then is it possible, in principle, to falsify that 
interpretation?

The answer, I believe, may be ``yes'', because none of the 
interpretations described above gives a quantitative account of the 
micro-macro divide -- that is, when do we, and when don't we, expect 
to see coherent superpositions?  To the extent that this becomes 
an experimentally answerable question, and to the extent that one 
can get corresponding predictions from the interpretations sketched 
above, one might be able to distinguish between different interpretive 
frameworks for an exact QM.  

\section{Theories Where Quantum  Mechanics is Modified}

Let us turn now to approaches based on premise (B), that QM is 
a very accurate approximation to a deeper level theory.  We will 
first discuss phenomenological approaches based on this premise, and 
then turn to attempts at a more fundamental theory. 

\subsection{Phenomenological Modifications:  Stochastic Models}

As we discussed in Subsect. 1.3, although the ``R'' process cannot 
be described by a deterministic unitary evolution, it is perfectly 
admissible for it to be described by a unitary evolution that differs 
for each individual measurement act, and in particular by a stochastic 
unitary evolution.  Considerable effort has gone over the past 
two decades into attempts to unify the ``U'' process and the ``R''
process into a single dynamical rule, by formulating 
 phenomenological modifications of the Schr\"odinger 
equation in which the ``individual'' is described by a stochastic 
unitary evolution of a pure state.  The physical motivation for 
such modifications is that if quantum theory is an approximation to 
physics at a deeper level, there could be small fluctuation or 
``Brownian motion'' corrections to this physics, which determine the 
outcomes for individual systems.

The natural mathematical language for formulating stochastic 
modifications of the Schr\"odinger equation is the It\^o stochastic 
calculus, which is basically a differential calculus version of the 
theory of Gaussian random variables. (For a clear exposition of the 
It\^o rules, see Gardiner [13].)  One introduces the stochastic 
It\^o differential $\D W_t$, which obeys the rules 
\begin{eqnarray}
(\D W_t)^2=\D t~,~~~\D W_t \D t=0~~;
\end{eqnarray}
thus $\D W_t$ is a fluctuating variable with magnitude 
$(\D t)^{1 \over 2}$, and as is familiar from the theory of path 
integrals, quantities of order $\D t$ are retained while those of 
order $(\D t)^{3\over 2}$ are dropped.  Let us now consider the 
following equivalent stochastic evolutions (introduced at various times, 
and in various forms, by Di\'osi; Ghirardi, Rimini, and Weber; Gisin; 
Hughston; Pearle; and Percival -- for references, see [14] and [15]).  
Letting $|z \rangle$ be a pure state, and $\rho = |z\rangle \langle z| 
/\langle z| z \rangle$ be the corresponding density matrix, we can 
write a stochastic pure state evolution 
\begin{eqnarray}
\D |z\rangle =[\alpha \D t + \beta \D W_t]|z \rangle~,\\
\alpha=-\I H -{1 \over 8} \sigma^2 [A-\langle A \rangle]^2 ~,~~~
\beta={1\over 2} \sigma [A-\langle A \rangle ]~,\\
A=A^{\dagger}~,~~~\langle A \rangle = \langle z|A|z \rangle /
\langle z| z\rangle ~~~,
\end{eqnarray}
or the equivalent [15] density matrix evolution 
\begin{eqnarray}
d\rho=-\I [H,\rho] \D t - {1\over 8} \sigma^2 [A,[A,\rho]] \D t
+{1 \over 2} \sigma [\rho,[\rho,A]] \D W_t~~~.
\end{eqnarray}
Here we have taken units with $\hbar=1$, 
$\sigma$ is a numerical parameter which governs the strength 
of the stochastic process that modifies the standard Schr\"odinger 
dynamics, and one can generalize the above equations by replacing 
$A \to A^j~,~~\D W_t \to \D W_t^j$ and including a sum over $j$ in 
each term.  Letting $E[\rho]$ denote the stochastic expectation of 
$\rho$ with respect to $\D W_t$ ( 
\emph{not} the same as the quantum expectation $\langle \rho\rangle$), 
the evolution of $\rho$ implies the following Lindblad type 
evolution of $E[\rho]$, 
\begin{eqnarray}
{\D E[\rho] \over \D t}= -\I [H,E[\rho]]
-{1 \over 8}\sigma^2 [A,[A,E[\rho]]]~~~.
\end{eqnarray} 

Let us now ask [15], when does this equation admit stationary 
solutions $E[\rho]_S$, for which 
\begin{eqnarray}
{\D E[\rho]_S \over \D t}=0= -\I [H,E[\rho]_S]
-{1 \over 8}\sigma^2 [A,[A,E[\rho]_S]]~~~?
\end{eqnarray} 
Multiplying by $E[\rho]_S$, taking the trace,  and using cyclic 
permutation under the trace, which implies that 
\begin{eqnarray} 
{\rm Tr} E[\rho]_S [H,E[\rho]_S]={\rm Tr} H[E[\rho]_S,E[\rho]_S]=0~~~,
\end{eqnarray}
we get the condition 
\begin{eqnarray}
0=-{1 \over 8} \sigma^2 {\rm Tr} E[\rho]_S[A,[A,E[\rho]_S]]\\
=-{1\over 8} \sigma^2 {\rm Tr}[A,E[\rho]_S][A,E[\rho]_S]^{\dagger} ~~~.
\end{eqnarray}
Since the argument of the final trace is positive semidefinite, it must 
vanish, and so we learn by reference to the evolution equation for 
$E[\rho]_S$ that we must have 
\begin{eqnarray}
[A,E[\rho]_S]=0~,~~~[H,E[\rho]_S]=0~~~,
\end{eqnarray}
in other words, a stationary value $E[\rho]_S$ must commute with both 
the Hamiltonian $H$ and with the operator $A$ which drives 
the dissipative process.  

Various cases are possible, depending on the choice of $A$:
\begin{itemize}
\item  For an \emph{energy driven process}, with $A=H$, the stationary 
value $E[\rho]_S$ can be any function of $H$.  One can then prove [15] 
\emph{with no approximations} that, when all energy eigenstates 
are nondegenerate, in the limit of large times, $\rho$ approaches 
an energy eigenstate projector $|e\rangle \langle e|$, with each 
such projector occuring as the outcome of the stochastic process 
with the corresponding probability $P_e ={\rm Tr} \rho(0) 
|e\rangle \langle e|$, with $\rho(0)$ the initial time density matrix. 
Correspondingly, the stochastic expectation of $\rho$, which is what 
we customarily term the density matrix, evolves from a pure state 
density matrix $E[\rho(0)]=\rho(0)$ to the mixed state density matrix 
$E[\rho]_S=\sum_e P_e |e \rangle \langle e|$.  Thus, for an energy 
driven process (or more generally, processes in which there 
are several  $A^j$ which all 
commute with the Hamiltonian), the QM probability rule 
is \emph{emergent} 
from the phenomenological stochastic dynamics.
\medskip\protect\newline 
As discussed in [14], if one assumes that the parameter $\sigma$ 
and the corresponding stochastic process originate from Planck scale 
physics, one gets the estimate (again in units with $\hbar$=1) 
$\sigma \sim M_{\rm Planck}^{-{1 \over 2}}$, which implies a 
characterstic state vector reduction time scale 
\begin{eqnarray}
t_R \sim \left(2.8 {\rm MeV} \over \Delta E \right)^2 {\rm sec}~~~,
\end{eqnarray}
with $\Delta E$ the energy dispersion of the initial state.  
An important question that remains to be answered is whether this
estimate gives a satisfactory phenomenology for state vector reduction 
in all cases, when the characteristic $\Delta E$ arising from 
environmental interaction effects is assumed.  That is,does the 
predicted micro-macro divide always occur in the right place?

\item For a \emph{localization process} (the Ghirardi, Pearle, Rimini
 form of the original Ghirardi, Rimini, Weber idea; for references 
see [14], [15]), one takes $A$ to be an operator that produces a 
Gaussian localization, or one uses multiple $A^j$ corresponding to many 
such localizations.  Since the kinetic term $\vec p^2/(2M)$ in the 
Hamiltonian $H$ does not commute with $\vec x$, and thus does not 
commute with a localizing operator $A$, there now is \emph{no} 
stationary limit unless the usual Schr\"odinger evolution term 
in the stochastic Schr\"odinger equation is neglected.  In this 
approximation, in which only the stochastic terms are kept, one gets 
similar results to those found in the energy driven case, with 
$H$ eigenstates replaced now by $A$ eigenstates.  
\medskip\protect\newline
For the localizing case, an important issue that remains to be 
addressed is whether the phenomenological theory can be made 
relativistic.  This may be a more severe problem than in the energy 
driven case 
because, while the Hamiltonian operator $H$ of Schr\"odinger 
dynamics  appears in a similar role in 
quantum field theory, the coordinate  $\vec x$ appears in quantum 
field theory as a label for degrees of freedom, 
rather than as an operator.   
\end{itemize}

\subsection{Fundamental Modifications}

Of course, even though the phenomenological stochastic Schr\"odinger 
equations discussed above give exactly (in the energy driven case) or 
approximately (in the localizing case) an emergent QM probability rule, 
there is still a probabilistic postulate in the form of the 
appearance of the It\^o differential $\D W_t$.  Since these equations 
have the characteristic form expected for the dynamics of open 
quantum systems, it is natural to ask whether they are simply the 
Brownian motion description of some underlying  dynamics.  
Specifically,  can one achieve a fully 
emergent probabilistic structure at the QM level from a pre-quantum 
dynamics that is not probabilistic?

Two approaches of this type have been discussed in the 
literature:   
\begin{itemize}
\item In [16] 't Hooft has proposed that quantum states are the 
equilibrium limit orbits or Poincar\`e cycles of an underlying 
chaotic, dissipative, deterministic theory.
\item In [17] we have proposed as a possible 
pre-quantum dynamics a ``generalized quantum'' or ``trace'' dynamics, 
obtained by setting up a generalized classical dynamics of noncommuting 
phase space variables $\{q_r\},\,\{p_r\}$ with no a priori commutativity 
properties beyond cyclic permutation inside a trace. One can show that,
with an approximation similar to assuming a large hierarchy between 
the pre-quantum and the QM energy scales, that by an equipartition 
argument the canonical commutation relations of QM are an emergent 
property of the statistical mechanics of this system.
\end{itemize}

Both of these proposed approaches to an emergent probability 
structure in QM are at present  programmatic, and  
significant open questions 
remain:  Can one construct an effective wave function and 
Schr\"odinger equation from the pre-quantum dynamics?  What do the 
leading fluctuation corrections look like, and are they the 
mechanism responsible for state vector reduction?
Can one use them to make 
contact with the phenomenological stochastic extensions of the 
Schr\"odinger equation discussed above?  Affirmative answers to these 
questions would yield the probabilistic structure of QM as an 
emergent phenomenon, in close analogy with the origins, in the 
underlying deterministic layer of molecular dynamics, of the 
probabilistic structure of statistical mechanics.   
Failure,  after sufficient effort, to construct 
such a pre-quantum underpinning for QM would support the view that 
QM is exact, in need perhaps only of a modified interpretation.

\section*{Acknowledgments}
This work was supported in part by the Department of Energy under 
Grant \#DE-FG02-90ER40542.

%

\end{document}